# VLSI Implementation of Cryptographic Algorithms & Techniques: A Literature Review

Favin Fernandes[1], Gauravi Dungarwal[2], Aishwariya Gaikwad[3], Ishan Kareliya[4], Swati Shilaskar[5]

Department of Electronics and Telecommunications, Vishwakarma Institute of Technology, Pune, India

[1]favin.fernandes18@vit.edu, [2] gauravi.dungarwal_18@vit.edu, [3]aishwariya.gaikwad19@vit.edu, [4]kareliya.ishan18@vit.edu, [5]swati.shilaskar@vit.edu

**Abstract**- **Through the years, the flow of Data and its transmission have increased tremendously and so has the security issues to it. Cryptography in recent years with the advancement of VLSI has led to its implementation of Encryption and Decryption techniques, where the process of translating and converting plaintext into cypher text and vice versa is made possible. In this paper, the review of various aspects of VLSI's implementation of encryption and decryption are covered. To systemize the material, the information about topics such as Private Key Encryption, Index Technique, Blowfish Algorithm, DNA cryptography, and many more are reviewed. Ultimately, with this review, the basic understanding of different VLSI techniques of Encryption and Decryption can be studied and implemented.**

## I. Introduction

In the modern era use of applications and websites have increased the use bank transactions, electronic mail, private messages etc., for these reasons secure communication is very essential. Secure communication requires a process of functions, which insures that the data is not accessed by unauthorized person over an unsecure medium. Thus, in order to build such a process; Cryptography is must.

Encryption and Decryption are the two functionalities of cryptography [1], the method that is designed to prevent unauthorized party/group access to data. In Cryptography the data is first encrypted into unrecognizable message to transfer the data and then its decrypted to its original data at the receiver side.

Encryption and Decryption are usually implemented using Algorithms that consists of key types [3], such as:

Symmetric Key – This algorithm uses the same cryptographic keys for both encryption of original data and decryption of the cipher text. (Encrypted original data)

Asymmetric Key – This algorithm uses 2 pairs of key for encryption, the First key is the public key i.e. available to any user while the second key i.e., the secret key is made available to the receiver. Both the key type uses the public key & Private key in order to encrypt the data.

Over the recent years the internet gave rise to many types of data, and the security of the data gave rise to Cryptography, which in turn gave rise to the implementation of VLSI for encryption and decryption using various techniques. In cryptographic applications, the data sent to a remote host are encrypted at the source machine using an encryption key then the encrypted data are sent to the destination machine, where it's decrypted to get the original data, thereby attacker won't have the encryption key which is required to get the original data.

Reprogrammable devices [2][4] like Field Programmable Gate Arrays (FPGAs) are used for hardware implementations of cryptographic algorithms. As FPGA devices progressed both in terms of resources and performance, the newest FPGAs provide solutions that are easily customizable for system connectivity, DSP, and data processing applications.

Implementing cryptographic algorithms [5] with VLSI has to take into its throughput, memory and power, similar to other VLSI designs. On top of this, the FPGA owner/designer also has to make the design resistant to physical attacks. In order to support the cryptography algorithms the following are needed: secure key storage, random number etc.

Further on we will discuss the different algorithms with the use of Random bit generation, DNA encryption and so on, implemented using VLSI to encrypt and decrypt images, files, data etc.

## II. A Review of Cryptographic Methodologies & Algorithms

### 1. VLSI Implementation of Text to Image Encryption Algorithm based on Private Key Encryption

This section conveys a method for text-to-image encryption and decryption using a private key [9]. The authors Anjali Suresh and Remya Ajai A S [6] proposed this and they tell about the cryptographic applications in this domain, the data sent to a remote host are encrypted at the source machine using an encryption key then the encrypted data are sent to the destination machine, where it's decrypted to get the original data, Field Programmable Gate Arrays are used for the hardware implementation of this algorithm. The implement of the encryption algorithm in FPGA is done using Xilinx ISE and is implemented on low cost Xilinx FPGA SPARTEN 3E.

A. TEXT TO IMAGE ENCRYPTION ALGORITHM

The text-to-image encryption (TTIE) algorithm in Fig.1 is shown below.

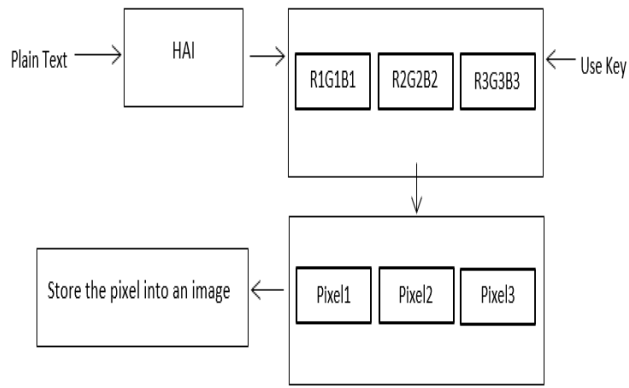

Fig.1 Encryption of Plaintext

Each letter in the text file is encrypted into three random integers say R for red, G for green, and B for blue where each random number ranging from 0 to 255. The three random numbers represent one pixel within the image file. This will make it difficult for hackers to attack the data transmitted over a network. This algorithm generates a random pixel R, G, and B values (calling them RGB values) for each letter. During this process the random numbers RGB for all letters are generated and used as keys for encryption, then use the key to transfer plain text into ciphertext. This key's generated at the receiver side, which is employed for decryption later. The result of this process is a two-dimensional array or matrix of pixels in which each pixel represents one letter.

B. DECRYPTION ALGORITHM:

The reverse algorithm in Fig.2 is shown below,

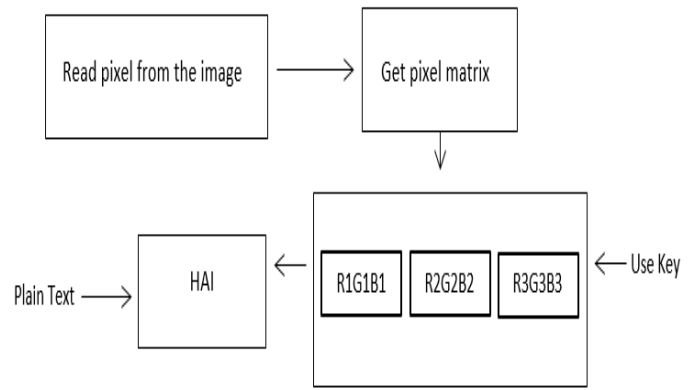

Fig.2 Decryption of Plaintext

The key generated is 24 bits for each character, thus making the length of key 64 bits for all 26 characters (A to Z). This longer key length makes this algorithm stronger or difficult to compute while considering all possible combinations. We have 26 letters and thus the permutations for 26 letters is $((256)^3)^{26}$. For an attacker to do all possible permutations will require many years which is not practical. This makes it stronger compared to other data encryption algorithms.

### 2. VLSI Implementation of Text to Image Encryption Algorithm based on Private Key Encryption

In this section, the authors Ammu S and Remya Ajai A S [14] discuss many aspects of security and many applications, such as secure commerce and payments to private communications and protecting passwords, here cryptography, and steganography comes into play.

FPGA implementation provides an intermediate solution between general-purpose processors (GPPs) and application specific integrated circuits (ASICs). FPGA configuring software makes use of the broad range of functionality supported by the reconfigurable device. So it has wider applicability than ASIC. FPGA provides a faster hardware solution than a GPP.

Encryption flowchart is shown in Fig.3:

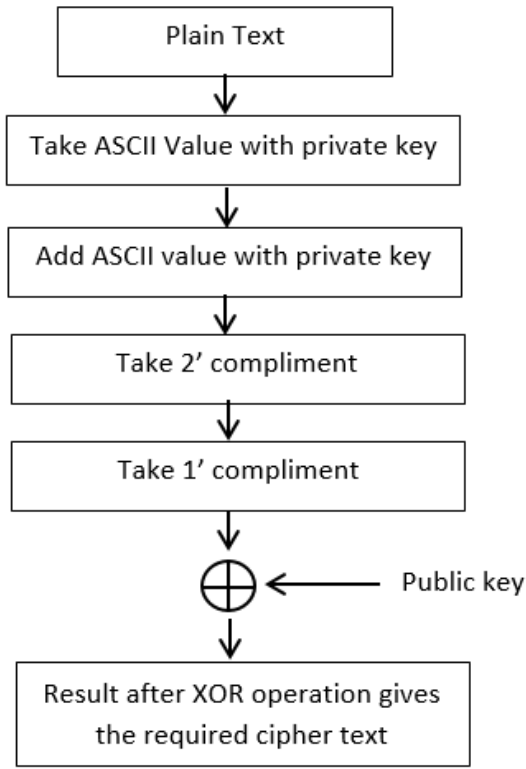

Fig.3 Encryption of Plaintext

The decryption of text:

The reverse algorithm in Fig.4 is shown below,

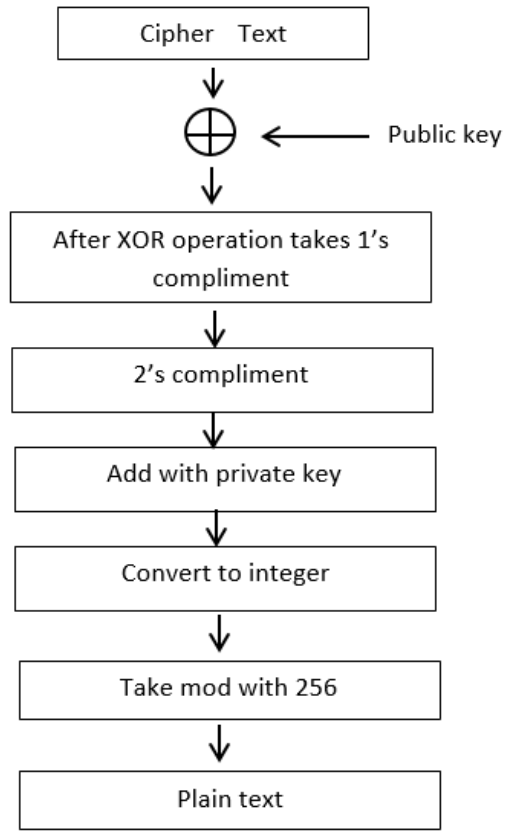

Fig.4 Decryption of Plaintext

Public key is 32 bit long, but the length of private key is dependent on the plain text. Characters presented are ranging from 0 -255, so the public key length should be maximum of (256*8 bit) i.e. 2048 bit. So there is ($2^{2048}$) possible combinations of keys are present. To decrypt a message attacker has to check ($2^{2048}$) combinations of keys. For an attacker to check all the possible combinations will require many years. So this algorithm is very secure.

**3. An Efficient VLSI Architecture for Data Encryption Standard and its FPGA Implementation**

A top-level view of proposed architecture by J. G. Pandey, Aanchal Gurawa, Heena Nehra, A. Karmakar [22] for the DES algorithm is shown below,

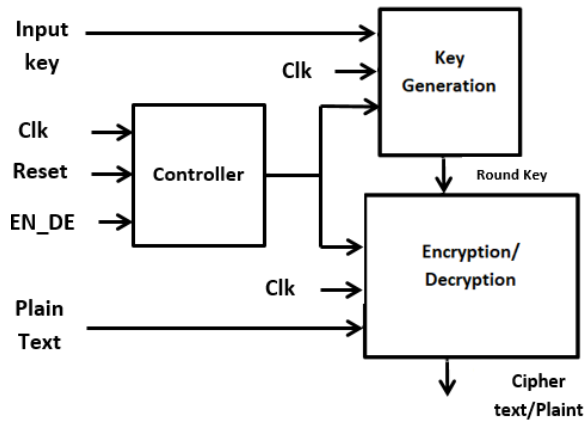

Fig.5 Block Diagram of DES Algorithm

In Fig.5, both the encryption and decryption operations use the same set of hardware building blocks. The three main components of the DES encryption/decryption computing scheme are key generation, controller, and an encryption/decryption engine. The key generation block is used to take 64-bit input keys and it generates sixteen, 48-bit round keys for the sixteen individual rounds. The main building blocks of the proposed architecture have been arranged in different subsections, which are described below.

A. Round Keys Generation Process:

In proposed DES algorithm, sixteen rounds are required to generate the sixteen round keys. To start the round key generation process, first, a 64-bit input key is provided to the permutation choice-1. The permutation block permutes the input key into 48-bit data, which are provided to two 28-bit blocks. After that, the two 28-bit data are circularly left-shifted by one or two bits, which depends upon the number of rounds as per the DES key generation algorithm. The encryption/decryption operation is managed by the controller.

B. Controller for the Encryption/Decryption Operation:

The controller is designed for providing necessary control signals to the proposed architecture. The proposed controller has six different states, which are shown in a form of a finite state machine (FSM). When the asynchronous reset input is at the logic '1' value, the machine resides in the Idle state and generates two control signals, both of these control signals are kept at the logic '0' level. After generating the above control signals, the machine waits for an external input signal, which is used to select the encryption/decryption operation. When the external signal is at logic '1', the machine enters into the Encrypt state otherwise the next state would be a Decrypt state.

C. Architecture for a Substitution Box:

The substitution operation consists of a set of eight substitution boxes (S-Boxes). Each S-Box accepts 6-bit input and it provides 4-bits output. The S-Boxes can be designed by using only a set of five multiplexers (MUXs). Using this approach, the proposed architecture for the realization of an S-Box using the MUX-based approach is done.

D. The Encryption/Decryption Operation:

The proposed architecture works for both encryptions as well as for decryptions, which depends upon the external input signal. When the signal external input signal is at logic '1', the controller resides in the Encrypt state and it asserts it to logic high. The architecture performs the encryption operation. Similarly, when the external input signal is at logic low in the Decrypt state, the architecture performs the decryption operation.

The operations are executed sixteen times and it is controlled by the various control signals generated by the controller. After completion of all the rounds, the registered output is available after nineteen clock cycles. Depending on the encryption/decryption operation, which is controlled by the external input signal, the generated output is ciphertext or plaintext.

## 4. VLSI Implementation of Hybrid Cryptography Algorithm Using LFSR Key

In this research, Shailaja Acholi and Krishnamurthy Ningappa [31] mentioned about hybrid cryptographic algorithm - Extended Tiny Encryption Algorithm (XTEA) combined with International Data Encryption Algorithm (IDEA) which was implemented for improving the security in real-time applications. The authors implemented this algorithm using Xilinx tool.

ID-XT-EA-LFSR algorithm:

The algorithm proposed by the authors employs five different algebraic operations: addition shift function, bitwise exclusive OR, addition modulo, and multiplication modulo. To increase the performance of both the process, the generation of the key using random numbers plays an important role.

The ID-XT-EA-LFSR algorithm includes six steps. In the first step, MATLAB is used to read the input image and then is converted into binary format. In the next step, the resulted binary value is converted to text format. This text format is given as input to Verilog. In the fourth step, LFSR is used to generate random numbers which is the key input to Verilog. The output of Verilog is then converted into text format. The resulted values are converted back to the pixels, and then in the final step, pixel values are converted into an image.

Hybrid cryptosystem consists of the initial stage of the XTEA architecture in the first half and in the second half, it includes IDEA architecture which is based on modulo multiplication.

8-bits data block is divided into 2-cycles of 4-bit each, and are represented as $v[0]$ and $v[1]$. The initial part is a permutation and the second part is sub-key generation. The sum function is used to select sub key block which is based on the 0th and 1st bits. In the encryption and decryption process, operations like XOR, addition, multiplication, and shift are used. The bits of $v[1]$ are shifted right and left by 5 and 4 respectively. The results of shift operations are XORed with each other. Then the value resulted by XORed is added to $v[1]$. R1 is the result. The delta value of 8 is added with the key. To produce r2, this output is XORed with r1. After this process, $v[0]$ is added with the value in r2 to produce r3. In the second stage of the proposed cryptosystem architecture, the initial value of $v[1]$ and the value of r3 is XORed which results in r5. The value in r5 is XNORed with the key which gives r6 as output. The values r5 and r6 are added and the result is XNORed with key. The output is represented by r7 and is added with r6 to give the output r8. The r8 is XORed with $v[1]$ and r3 results into V0 and V1 respectively. To produce the encrypted results V0 and V1 are concatenated. Finally, to get the decrypted results, the encrypted value is given to the $v[0]$ and $v[1]$.

LFSR generates random numbers that can be used as keys to transmit ciphers. LFSR is used as a series of flip-flops within the FPGA platform. The number of taps from the chain of the switch register is used for the XOR / XNOR gate. The output of this gate is used as a response at the beginning of the shift register chain, hence the response to the LFSR. When LFSR is operational, the pattern is produced by the flip flops individually by a pseudo-random number. Verilog creates a 4-bit LFSR key. It uses polynomials to make the maximum length of LFSR in a single width.

The proposed ID-XT-EA-LFSR algorithm was implemented in the FPGA platform. This platform is best suited for VLSI due to its flexibility, low power, and high compatibility compared to the ASIC platform.

## 5. Implementation of AES Algorithm in UART Module for Secured Data Transfer

Advanced Encryption Standard (AES) algorithm in Universal Asynchronous Receiver Transmitter (UART) module for secure transfer of data rate. The architecture implements AES-128 algorithm [32] that encrypts the data before transmission through UART transmitter and decrypts after receiving the data at UART receiver. The design has a clock pulse generator circuit which provides the different clock frequencies to different sub modules present. The complete design is described in Verilog Hardware Description Language (HDL) and is functionally verified using Xilinx ISE 9.1i software.

The authors had proposed an FPGA based approach [33], Encryption Module consists of sub parts, two of them are 128-bit shift register and one is AES encryption module. The first serial in parallel out shift register stores the data bytes in particular clock cycle and sends them to the AES-128 Encryption module for encryption process. The 128-bit encrypted data is transferred to the UART transmitter through parallel in serial out shift register. This AES Encryption module does the encryption operation on 128bit of data using AES algorithm which is a symmetric block cipher that processes data blocks of 128 bits using three different cipher key lengths which are 128, 192 &256 bits.

UART transmitter module i.e. UART_Tx that frames the 8-bit word coming from AES-128 encryption unit with a START bit is logic 0 at the beginning, and a STOP bit logic 1 at the end of the word and sends the framing information in a serial from the Least Significant Bit (LSB) to the Most Significant Bit (MSB). The architecture of the transmitter will consist of a controller, a data register (XMT_datareg), a data shift register (XMT_shftreg) and a status register (bit count) to count the bits that are transmitted. Load_XMT_datareg signal is asserted to indicate that XMT_datareg now contains the data bus value and that is now transferred to the internal shift register that is XMT_shftreg.

## 6. Implementation of Speech Encryption and Decryption using Advanced Encryption Standard

Comparing analog and digital signal the use of digital speech has become widespread and has advantages over analog voice signal This paper [34] presents design concept which includes all the operations to be performed on a specific chip. The design is intended to be implemented as an ASIC.

Serial Interface Design:

Encryption and decryption process continues till the end of data transmission [35] takes place. The end to end data transfer is shown in Fig below UART (Universal Asynchronous Receiver Transmitter) system was used for serial data transfer from computer to FPGA, FPGA to FPGA and FPGA to computer. RS232 a serial communication standard has been used for data transfer. The speed of the RS232 link is determined by its baud rate. The baud rate is the bit rate of the communication link in bits per second (bps). Both transmitting and receiving FPGA boards must communicate at same baud rate. In Fig.7, the UART control module that is designed contains receive module, transmit module, baud rate generator, receive data register and transmit data register.

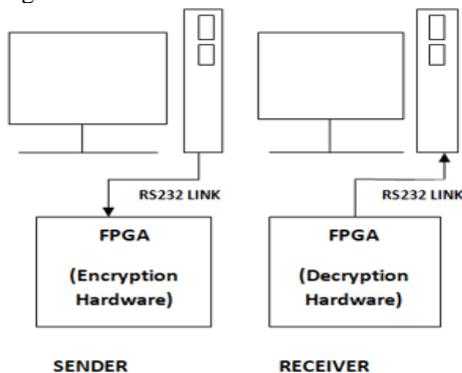

Fig 7. Control Module

The voice which is recorded in computer using MATLAB is shown [36]. The voice is sampled at the rate of 8000 samples per second. The recorded voice is of one second duration and so the total number of samples is 8000. The amplitude level of each sample is digitized and is of 8-bit length. The proposed architecture is synthesized using Verilog HDL (Hardware Description Language) in Xilinx ISE Design Suite. The proposed design has been implemented on Spartan 3E FPGA (XC3S500E) boards. MATLAB is used for translation between voice signal and digital data. The real time secured offline voice communication between two computers which is the most fascinating part of this design has been successfully implemented. Simulation results of encryption module and decryption module are shown. [36]

## 7. DNA cryptography

DNA cryptography is a new method providing high security based on DNA nucleotides bases A-Adenine, C-Cytosine, G-Guanine and T-Thymine. In this proposed model P.Vinotha and Deepa Jose used Polymerase Chain Reaction encoding technique[40] in which the image to be coded is flanked between primer keys.

### A. DNA ENCODING

The image is converted to binary values based on the pixel value of the image. Binary code is converted into DNA code shown in Table.1 by providing the bases of DNA in alphabetical order (A, T, C, and G). DNA sequences are designed for the whole picture of high pixel values 0-255. [38]

| BINARY CODE | DNA CODE |
|---|---|
| 00 | A |
| 01 | C |
| 10 | G |
| 11 | T |

Table.1 DNA ENCODING

The natural structure of DNA is expressed in the form of PCR Encoding technique. The required information to be coded is between F 'Primer and R' Primer sequence. Primer sequence is referred to as genetic marking in a sequence marked with different colours in a natural structure. Here, using PRBS the Primer key is generated and is considered as OTP shared between sender and receiver.

### B. ENCRYPTION PROCESS

The encoding process discussed here uses DNA based Polymerase Chain Reaction encoding. The process is summarized below in Fig.8.

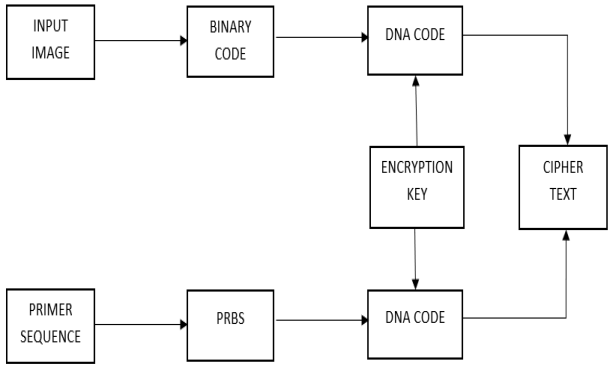

Fig.8 Block diagram of encrypting an image

The encryption steps are as follows
• Provide input data: Image
• The image is converted to binary code using MATLAB
• Each pixel value of an image is converted into 8-bit binary code.
• Hamming Code: Error correction code used to detect and correct minor errors it is possible when data is transferred or stored. The transfer channel should add more data in data to be transferred.
• DNA code: Two-way A-'00 ', T-'01', C-'10 ', G-'11' of 8-bit form.
10 00 11 01 (141) → GATC
DNA code is converted into ciphertext by considering the nucleotides of the fourth base providing a key combination of four bases.
• Final text is the same part containing full values, alphabets, specials letters, and symbols in the DNA data format.
• Primer key made by PRBS in the form of DNA sequences is converted into text and embedded text messages are inserted between these primer keys.
• Ciphertext is converted to an encrypted image using MATLAB.
The initial encryption key and sequence are shared between sender and recipient at
a type of microdots or other communication systems.

## C. KEY GENERATION (OTP) USING PRBS

Primer keys are generated by the process of producing an OTP key (One Time Pad) with the help Pseudo-Random Sequence Generator (PRBS) [40]. The main function of PRBS register to change direct response. Contains a group of flip-flops in the series with feedback loop. The primer sequence is given as a seed in the LFSR (Linear Feedback Shift Register) randomly produced Binary sequence used as primer key. Binary sequence is converted into DNA code. The primer key is used only once in one image.

## D. ENCRYPTION KEY

In this method, the ciphertext was created by insert alphabets, symbols, and special characters into the base of the DNA codon of length 4. The possible combination of DNA codons is 256 to increase the key size. These are 256 the combination of keys and their DNA sequence serve as the encryption key between the sender and the recipient.
Table.2 shows the combination of keys and their DNA sequence.

| A = AAAA | K = AATG | U = GGTA |
|---|---|---|
| o = TTTT | y = TCAA | p = ACTT |
| # = GGGG | r = CTAG | @ = GATT |
| u = ATCC | H = CTAG | ' = TCAA |
| e = AACA | z = ATGC | ? = GTAA |
| t = TTAC | T = CCGA | % = GGAC |
| < = AGCC | s = GTAC | Y = CCTA |

Table.2 Encryption key

In this way, the time for encryption and complexity is reduced. The size of The key is relatively large and the text to be changed is reduced. Deciphering the image is not possible with the primer sequence information. The method is very safe and prevents the enemy from deciphering it.

## E. SECURITY ANNALYSIS

1. Key Used in DNA Based AES Encryption

The key length used for AES-based encryption by AES is 64 codons of third base DNA [39]. A combination of keys that can be $4^{64}$. The chances of a successful guess are the attacker is given
$$P \text{ (SG of Key)} = 1 / (4^{64})$$

The total probability of successful guess by an attacker is
$$P \text{ (SG of Key)} = 1 / (1.63 * 10^{8} * 16! * 4 * 4^{64})$$

2. Key Used in DNA Based PCR Encoding Technique

In the proposed model the maximum length used in the PCR DNA coding techniques approx. 256 DNA bases. It produces a combination of about $4^{256}$ keys. Chances are a successful guess by the attacker is given
$$P \text{ (SG of Key)} = 1 / (4^{256})$$

The total probability of successful guess by an attacker is
P (SG of Key) = 1 / (1.63 * 10^8 * 16! * 4 * 4^256)

The length of the four nucleotide sequences increases the base of the key up to 256. There are 4,29,49,67,296 possible combinations of DNA sequences. The chance for Image clarification is possible with the primer key and encryption key.

## 8. Advance Encryption Algorithm using Index Technique

M. M. Abdelwahab and A. J. Alzubaidi implemented the Index Technique [42], a fixed advance encryption algorithm in which storing keys are applied for fixed encryption operation, the person who owned the FPGA device which contains the algorithm can encrypt/decrypt any information.
This technique is optimized because it has a short path of encryption.

The implementation consists of two fixed rounds of encryption with two different key which is not known for users.
One important issue is that the used area of the chip is an important factor that affect performance, in other words increasing the number of routing will decrease the performance speed.

In short, we'll discuss the Algorithms for the Indexing technique

Encryption Algorithm: In this algorithm, users are not required to have the key because it is stored automatically inside the algorithm.
Mathematically the algorithm is as follows:

1- Dividing the plaintext into 16 bytes.

2- At this stage, each Aij element is added with a corresponding element of an internal key. This defines how the keys are added automatically in a particular transposition and show that users are not involved in this operation. The keys are stored on the LUTs inside the FPGA. In the final operation, mix the output cipher using the shift technique

3- The output cipher of stage 2 rearranged by shifting its elements.

Decryption algorithm: The same three steps illustrated in the previous encryption algorithm repeated in order to obtain the original message, this operation starts by dividing the input cipher into bytes followed by step 3 using reverse transposition then adding the algorithm keys.

The keys are stored in the algorithm and applied in two rounds of encryption and decryption
The results are carried out on several types of FPGA and shown below.
In Table.3 the speed grade table is shown below, the targeted devices by the authors are spartan3 and vertix5:

Time Report

|  | Spartan3 Speed grade -5 | Vertix5 Speed grade -5 |
| --- | --- | --- |
| Maximum frequency | 194.727MHz | 419.674MHZ |
| Minimum input arrival time before clock | 3.905ns | 1.130ns |
| Maximum output required time after clock | 7.471ns | 3.130ns |
| Minimum Period | 5.135ns | 2.383ns |
| Maximum Combinational path delay | 9.750ns | 3.393ns |
| Throughput MBPS | 853.245 | 853.746 |

Table.3 Time Report of FPGA's

## III. Conclusion

Cryptography aims to provide high data security and plays an important role in data exchange. In this review, various Cryptographic Algorithms were covered with brief information about the working, design, implementation using VLSI and how these algorithms are used in cryptography for security purpose. These algorithms are developed to achieve security goals like authentication, confidentiality, integrity. This review was aimed to cover the algorithm and its functioning to help researchers to attain knowledge of a particular technique to implement in his/her project or research. Utilizing FPGA's in the near future will be highly appreciated for Cryptographic Algorithm and further on, study needs to be taken on different FPGA attacks and its safety.